\documentclass[prd,preprint,superscriptaddress]{revtex4}
\usepackage{amssymb}
\usepackage{revsymb}
\usepackage{graphicx}

\newcommand{\ds}{\displaystyle}
\newcommand{\be}{\begin{equation}}
\newcommand{\en}{\end{equation}}
\newcommand{\bea}{\begin{eqnarray}}
\newcommand{\ena}{\end{eqnarray}}

\newcommand{\ov}{\overline}
\begin{document}
\title{Late Universe expansion dominated by domain walls and dissipative dark matter}
\author{Sergio del Campo\footnote{E-mail address: sdelcamp@ucv.cl}}
\affiliation{Instituto de F\'\i sica, Pontificia Universidad
Cat\'olica de Valpara\'{\i}so, Av. Brasil 2950, Casilla 4059,
Valpara\'\i so, Chile}
\author{Ram\'{o}n Herrera\footnote{E-mail address:
ramon.herrera.a@mail.ucv.cl}}
\affiliation{Instituto de F\'\i sica, Pontificia Universidad
Cat\'olica de Valpara\'{\i}so, Av. Brasil 2950, Casilla 4059,
Valpara\'\i so, Chile}
\author{Diego Pav\'{o}n\footnote{E-mail address: diego.pavon@uab.es}}
\affiliation{Departamento de F\'{\i}sica, Universidad Aut\'onoma de Barcelona,
 08193 Bellaterra (Barcelona), Spain}
\begin{abstract}
In this paper we show that a universe dominated by two components, namely
domain walls and dissipative dark matter such that the former
slowly decays into the latter may drive power law
cosmological acceleration consistent with the high redshift
supernovae data, while the ratio between the energy
densities of both components remains fixed at late
times thus solving the coincidence problem of present 
acceleration. Likewise, we estimate the aforesaid ratio
at early times.
\end{abstract}
\maketitle
\section{Introduction}
In the last years measurements of the luminosity--redshift of type Ia
supernovae have provided growing evidence that our  Universe has
recently entered a phase of accelerated expansion \cite{supernova,knop}.
In Einstein gravity accelerated expansion of a spatially flat
universe requires that the cosmological dynamics be dominated by some
exotic component with a negative pressure high enough to violate the
strong energy condition.

The position of the first acoustic peak of the CMB as witnessed by
the WMAP satellite \cite{wmap} combined with large scale observations
of mass distribution suggests that this may be the case as the spatial
geometry is nearly flat while the mass density in clustered matter is
roughly half the critical density \cite{Peacock}. This is also
consistent with studies of the baryon fraction in galaxy clusters
\cite{White} together with big bang nucleosynthesis calculations
\cite{Schramm}. Therefore, some additional non--clustered form of
energy (dark energy) is needed to attain critical density.

The precise nature of the dark energy component has become a subject
of intense debate \cite{iap}. Dark energy candidates include the
cosmological constant (or vacuum energy), quintessence (a nearly
spatially homogeneous but time-dependent scalar field)
\cite{Caldwell}, tachyon field \cite{tachyon}, Chaplygin gas
\cite{chaplygin}, solid  dark energy \cite{Bucher}, a dense
network of  low-tension domain walls \cite{Hill,Friedland},
among others.

In introducing dark energy as a novel component most of these candidates
face the embarrasing question: ``Why are the energy densities of both
components of precisely the same order today?" This is the so--called
``coincidence problem" \cite{coincidence}.

The aim is this paper is to study a model of late acceleration
dominated by two components, namely a strongly frustrated network
of domain walls and dissipative dark matter. It is usually assumed
that domain walls and dark matter evolve independently.  An
interesting alternative is to consider that the evolution of the
domain walls  is slowed down by their interaction with the dark
matter fluid \cite{Massarotti}. In that model the average wall
velocity is determined by the ratio of wall and dark matter energy
densities.  In this respect, it seems worthwhile to explore a
model featuring a coupling between both components. Here, we
assume that the two components do not conserve separately but the
domain walls slowly decay in dark matter. This process could be
realized by a broken symmetry that gives domain walls and then the
symmetry could be restored and the walls would slowly decay
\cite{Private}. Here, by slowly we mean that
$\dot{\rho}_{dw}/\rho_{dw}<\,H$, where $\rho_{dw}$ is the energy
density associated to the domain walls (subscript $dw$),
$H=\dot{a}/a$ is the Hubble factor, $a(t)$ is the scale factor of
the Friedmann--Robertson--Walker metric and the over dots denote
derivative with respect to cosmic time. In this case the energy in
the walls would have to go into dark matter particles. As it 
turns out, this scenario is compatible with a stage of 
accelerated expansion governed by a near power law scale 
factor consistent with the high redshift supernovae data. In
addition, the ratio between the energy densities
$\rho_{m}/\rho_{dw}$ stays constant at late times. This solves the
coincidence problem that afflicts many approaches to late
acceleration without recourse to fine tuning. 

Domain walls are ubiquitous in field theory models with spontaneously
broken discrete symmetries \cite{Vilenkin} and they appear to be
compatible with the observations of large scale structure
\cite{Fabris}, and the CMB anisotropy spectrum provided that
the domain walls network is strongly frustrated \cite{Friedland}.
In such a case its equation of state parameter $w=p_{dw}/\rho_{dw}$
may vary with expansion according to
\\
\[
w = - \frac{2}{3}+ \frac{\alpha}{3} \, \frac{d \ln \, \eta}{d \ln \, a},
\]
\\
where $\eta$ the conformal time, $dt = a(t) d \eta$ and
$\alpha$($\leq 1$) a constant \cite{Friedland}. Further, Conversi
{\it et al.} \cite{conversi} have recently shown that even if the
network is not frustrated, no conflict with the CMB anisotropies
arises provided the reduced Hubble parameter is less than $0.65$
(as some experiments seem to indicate \cite{birkinshaw}) and the
matter density parameter lies above $0.35$ (as the case might be,
see Ref. \cite{blanchard}). In section II we explore the possibility 
that domain walls may act as dark energy in late expansion in such a
way that the coincidence problem may be solved.
This analysis extends previous studies of two interacting components
-quintessence and dark matter-, in which a constant stable ratio
arises between them at late times due to their mutual interaction
\cite{plb}, to a universe dominated by domain walls and dark matter.
Obviouly, the ratio $\rho_{m}/\rho_{dm}$ cannot have stayed constant
during the whole Universe history, otherwise the nucleosynthesis, CMB 
and cosmic structure scenarios would have been very different. This 
is why section III focus on the evolution of that ratio and evaluates 
it at the time of domain walls formation. Section V calculates the 
statefinder parameters. Finally, section V summarizes our findings.

\section{Domain walls-dark matter interaction}
We consider a spatially flat Friedmann--Lemaitre--Robertson--Walker
universe dominated by  two interacting components, a strongly
frustrated network of domain walls and dark matter (subscript $m$).
In the following we shall assume that the domain walls behave as a
fluid as a whole.  Consequently, the stress--energy tensor of the
cosmic medium takes up the hydrodynamic form and the corresponding
Friedmann equation is
\\
\be
\ds H^{2} = \frac{\kappa}{3} \rho \qquad \quad (\kappa \equiv 8 \pi G),
\label{Friedmann}
\en
\\
where $\rho = \rho_{m} + \rho_{dw}$ is the total energy density.

Since by assumption both components do not conserve separately
a source/loss term (say, $Q$) must enter the energy
balances
\\
\be
\ds \dot{\rho_m}+3H(\rho_m+p_m+\pi_{m})= Q,
\label{ecM}
\en
\\
and
\be
\ds \dot{\rho}_{dw}+3H(\rho_{dw}+p_{dw})=-Q,
\label{ecDW}
\en
\\
where $\pi_{m}$ stands for the dissipative pressure in the dark
matter fluid.  Barring superfluids, the dissipative pressure arises
rather naturally in all fluids in Nature \cite{landau,batchelor}
whereby we have no a priori reason to ignore it
(unless we knew that dark matter were in a superfluid state, something 
that does not seem to be the case).  Likewise, a comparatively
high negative pressure may arise within the dark matter fluid by the
decay of dark matter particles into dark particles \cite{pep}. The
decay of dark particles in the halo of galaxies has been considered
by different authors to explain the absence of steep cusp in the
halos of the galaxies -see \cite{salcedo} and references therein.
Further, the development of density inhomogeneities has been pointed
out as a source of negative pressure in the cosmic fluid
\cite{dominik}.

We are interested in studying the dynamics of the ratio $\rho_m/\rho_{dw}$
\\
\be
\ds \frac{d}{dt}\left[\frac{\rho_m}{\rho_{dw}}\right]=\frac{\rho_m}
{\rho_{dw}}\left(\frac{\dot{\rho}_m}{\rho_m}-
\frac{\dot{\rho}_{dw}}{\rho_{dw}}\right).
\label{din}
\en
\\
By virtue of equations (\ref{ecM}) and (\ref{ecDW}) we can write

\be \ds\frac{d}{dt}\left[\frac{\rho_m}{\rho_{dw}}\right]=
3H\frac{\rho_m}{\rho_{dw}} \left(w +
\frac{\rho}{\rho_m\,\rho_{dw}} \, \frac{Q}{3H} -
\frac{\pi_{m}}{\rho_m} \right)\, , 
\label{wecan} 
\en
\\
where for simplicity we assumed that the matter 
fluid behaves as pressureless dust.

A stationary stage, i.e., $d(\rho_m/\rho_{dw})/dt = 0$ is
achieved for
\\
\be
\ds Q=-3H\frac{w}{1+r_{0}}\rho_{dw}\left[r_{0}-\frac{\pi_{m}}
{w \, \rho_{dw}} \right],
\label{deltaDW}
\en
\\
where $r_{0}$ stands for the stationary value of the ratio 
$\rho_{m}/\rho_{dw}$ at present time.

To study the stability of this stationary solution against small
perturbations we introduce the ansatz
$\frac{\rho_m}{\rho_{dw}}= r_{0} + \epsilon $
into Eq.(\ref{din}) to get
\\
\be
\ds\dot{\epsilon}=3H\left[w + \frac{\rho}{\rho_{dw}\, \rho_m} \,
\frac{Q}{3H}- \frac{\pi_{m}}{\rho_m}\right] (r_{0}+\epsilon).
\en
\\
Following \cite{ladw} we assume that the interaction and the
dissipative pressure are related to the total energy density by
$Q = 3 c^{2} H \rho$, and $\pi_{m} = -b^{2} \rho$, where $c^{2}$
and $b^{2}$ are positive--definite constants. (It should be
borne in mind that $\pi_{m}$ is always negative for expanding
fluids \cite{landau}). Thus, up to first order in $\epsilon$
one obtains
\\
\be
\ds\dot{\epsilon}=3H\left[\frac{c^2(r_{0}^{2}\,-\,1)\,-\,b^2}{r_{0}}  \right]\epsilon.
\label{firstorder}
\en
\\
Therefore the stationary solution will be stable for
\\
\be
\ds\,r_{0}<\sqrt{ 1\,+\,\frac{b^2}{c^2}},
\label{menorr}
\en
\\
which is compatible with the observation that $\rho_{m}/\rho_{dm} < 1$
at present.

From (\ref{deltaDW}) it follows that
\\
\be
\ds\,c^2= - \frac{r_{0}}{(1+r_{0})^2}\left[w + b^{2} \, \frac{1+r_{0}}{r_{0}} \right].
\label{c2}
\en
\\
This sets on the dissipative parameter the upper bound $b^{2}< -r_{0}w/(1+r_{0})$ 
since as noted above $c^{2}$ must be positive.

From Eqs.(\ref{ecM}), (\ref{ecDW}) and (\ref{deltaDW}) along with
the assumption that $w$ varies slowly (or that it is piecewise
constant) the dependence of the energy densities on the scale
factor at late times is readily found
\\
\be
\ds\rho_{m}\propto a^{-\beta}, \qquad \rho_{dw}\propto a^{-\beta},
\label{abeta}
\en
\\
where
\\
\be
\ds\beta=3\left[\frac{w}{1+r_{0}}+1 -b^{2}\right].
\label{beta}
\en

\begin{figure}[tbp]
\includegraphics*[angle=0,scale=0.5]{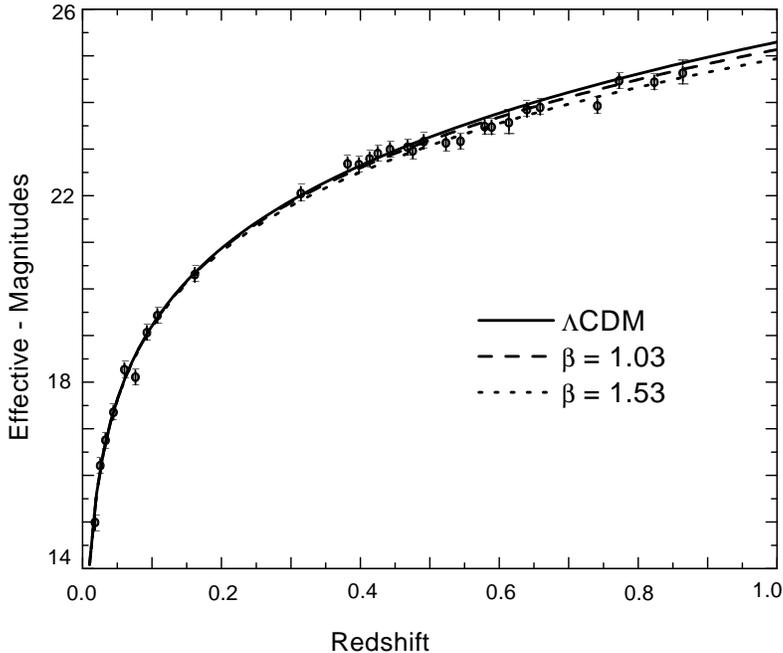}
\caption{{} Effective magnitude vs redshift for the model
discussed in the paper. To plot the graphs the expression
$m_{B}^{eff} = M_{Bf}+ 5 \log (H_{0}d_{L})$, with $M_{Bf} = -3.9$
and $d_{L} = (1+z) \int^{z}_{0}H^{-1}(z') dz'$, was used.
For comparison, we include the prediction of the $\Lambda$CDM 
model with $\rho_{m} =0.3$. The data were borrowed from 
figure 6 of Ref. \cite{knop}.} 
\label{snia}
\end{figure}

Combining Eq. (\ref{abeta}) with  Friedmann's equation (\ref{Friedmann})
we get
\\
\be
\ds a\propto \,t^{2/\beta} ,
\label{scalefac}
\en
\\
whereby accelerated expansion will occur for $\beta < 2$.
\\
When the Universe is dominated by just one component with constant
equation of state parameter $w_{*}$, the scale factor varies as
$t^{2/3(1+w_{*})}$. Therefore, in our case, the effective
parameter is
\\
\[
w_{eff} = \frac{w}{1+r_{0}} - b^{2}.
\]
\\
For the sake of numerical evaluation let us consider $r_{0} = 3/7$ 
-see e.g. Ref. \cite{iap}.  For $w = -0.5$ and $b^{2} = 0.14$, one
follows $w_{eff} = -0.49$, whence $\beta =1.53$; while for $w
= -0.66$ and $b^{2} = 0.19$, one obtains $w_{eff} = -0.66$,
whence $\beta = 1.03$.  These $w_{eff}$ figures are only
marginally compatible with analyses that tend to push $w_{eff}$
closer and closer to the cosmological constant value and even
beyond it \cite{iap,alessandro}. However, it should be noted
that, as observed by  Maor {\it et al.} \cite{maor}, these
analysis are based on combining luminosity--redshift measurements with 
models that consider the equation of state parameter of the dark energy
$w_{x}$ constant which automatically introduces a bias that favors
larger negatives values of $w_{x}$ than if it varied with
expansion. On the other hand, as Fig. \ref{snia} shows the
agreement of our model with the supernovae data of Knop {\it et
al.} \cite{knop} is excellent.

At this stage, one may wonder why at all one needs a coupling between
both components and a dissipative bulk pressure in the dark matter if
the the most negative value for $w_{eff}$ essentially coincides with
the one provided by domain walls ($-2/3$). The answer is that these
ingredients are key to guarantee that the ratio $\rho_{m}/\rho_{dw}$
will remain constant at late times. In our view, models of late
acceleration that fail to address the coincidence problem 
are not very useful.

\section{The ratio $\rho_{m}/\rho_{dw}$ at early times}
\begin{figure}[tbp]
\includegraphics*[angle=0,scale=0.5]{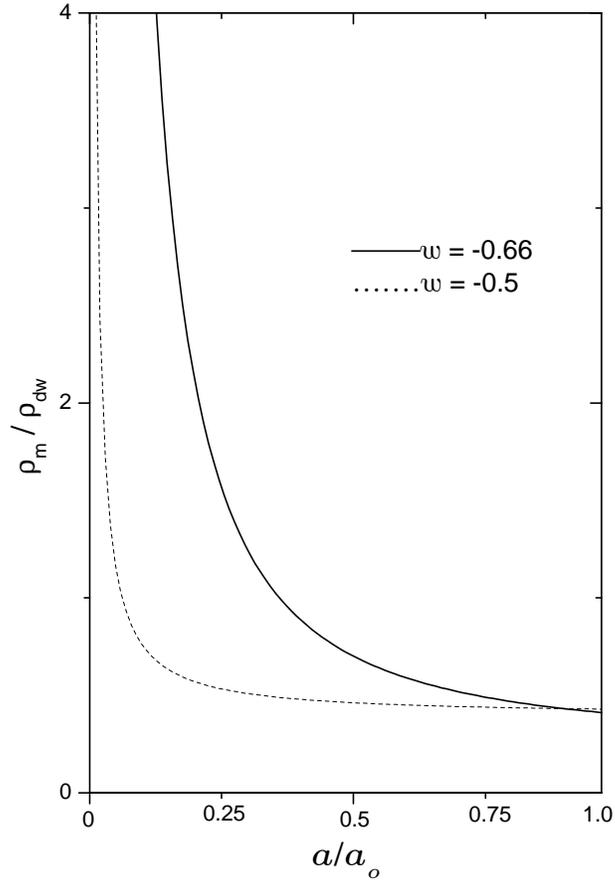}
\caption{{} 
The ratio between energy densities $\rho_{m}/\rho_{dm}$ decreases with expansion up to 
an asymptotic value at the present epoch. In drawing both curves we assumed
$b^{2} = 0.14$ and $c^{2}= 0.04$.} 
\label{evolr}
\end{figure}

Obviously, at the time of domain walls formation (about $10^{-11}$
seconds after the big bang \cite{em}) the ratio between  the
energy densities must have been very different from its present 
value, $r_{0}$. In this connection it is mandatory to study the 
evolution of $r \equiv \rho_{m}/\rho_{dm}$ with the Universe 
expansion if one wishes to evaluate $r$ at that epoch. 
Here we shall follow a procedure parallel to that of 
Ref.\cite{ladw}. The evolution equation for $r$ is 
given by Eq.(\ref{wecan}), with $Q = 3 c^{2} H \rho$ 
and $\pi_{m} = -b^{2} \rho$. Its stationary 
points,    
\\
\be
 r^{\pm}=-1-\left(\frac{b^{2}+w}{2c^2}\right)\,\pm\,\sqrt{\Pi}\,
\label{rmasmenos},
\en
\\
 where
\\
\be
\Pi=\frac{w^{2}}{4c^{4}}+\frac{w}{c^{2}}\left(\frac{b^2}{2c^2}+1\right)+\frac{b^4}{4c^4}\, ,
\en
\\
follow from setting  the right hand side of Eq.(\ref{wecan}) to zero with $w =$ 
constant. Notice that $\Pi$ determines the difference between the stationary
solutions $r^{+}-r^{-}=2\sqrt{\Pi}$, and that their sum is
$r^{+}+r^{-}=-2-(b^2+w)c^{-2}$ . To these relations the 
constraint
\\
\be 
r^{+}r^{-}=1+\frac{b^{2}}{c^{2}} 
\label{rrbc}
\en
\\
is to be added. The stability of the stationary points is determined by the sign of
$[\partial \dot{r}/\partial{r}]_{r^{\pm}}$. It can be seen that
that $r^+$ is unstable and $r^-$ is asymptotically stable. 
This tells us that the solution of Eq.(\ref{wecan}) 
evolves from $r^{+}$ to $r^{-}$ 
with $r^{+}>r^{-}$ -see Fig.(\ref{evolr})
In order to describe  the transition from the decelerated phase to the accelerated 
one we resort to the deceleration parameter $q=-\ddot{a}a/\dot{a}^{2}$. Before
domain walls dominance the Universe is decelerating, $q> 0$, and later changes to
$q<0$. In virtue of the Friedmann equation (\ref{Friedmann}) and the evolution
equation for the Hubble factor
\\
\be \dot{H} = - \frac{\kappa}{2}[\rho_{m}+(1+w)\rho_{dw} +
\pi_{m}]\,\label{dotH},
 \en
\\
we can write
\\
\be q=\frac{3}{2}\left[1+\frac{w}{1+r}-b^2\right]-1\,
\label{qq}.
\en
\\
It is worthy of note that the condition $q^{+}=q(r^+)>0>q(r^-)=q^{-}$, implies
the constraint 
\\
\be
\frac{w}{1+r^{-}}+\frac{1}{3}\,<\,b^2\,<\,\frac{w}{1+r^{+}}+\frac{1}{3}\,.
\en
\\
Likewise, it is convenient, for later purpose,  to estimate $c^{2}$ in terms 
of $r^{+}$ and $r^{-}$. From  Eqs.(\ref{rmasmenos}), (\ref{qq}) 
and the constraint (\ref{rrbc}), we readily obtain
\\
\be
c^{2}=\frac{1-2q^{-}}{3r^+(1+r^{-})}\, ,
\label{pri}
\en
\\
in agreement with the previous statement that $c^2$ 
ought to be a positive-definite quantity. \\

The condition $q(r_{ac})=0$ yields the density 
ratio at the beginning of the acceleration 
\\
\be 
r_{ac}=\frac{-3w+3b^{2}-1}{1-3b^{2}}\, . 
\en
\\
Using the restriction $3b^{2}<1$ we get $r_{ac}\geq -(1+3w)$. Obviously
$r^{+}>r_{ac}>r^{-}$, as can be checked. 

The observational constraints of type Ia (SNeIa) supernovae may 
be used to determine the values of $c^2$ and $b^{2}$. For each 
supernova, its redshift $z_{i}$, dispersion $\sigma_{i}$ and  the
corrected magnitude $m_{i}$, were computed, resulting $\sigma\simeq\,0$ 
as the most likely value of $\sigma$  \cite{supernova}. We may 
tentatively assume that for $\sigma=0$ the Universe is settled in an
asymptotic energy density ratio $r^{-}\simeq\,r_{0}$,
with constant deceleration parameter $q^{-}$. On
the order hand, for power-law expansion given by
Eq.(\ref{scalefac}), we find $d_{L} \propto(1+z)[(1+z)^{\delta}-1]$,
with $\delta=-q^{-}$. Thus,  we get $0<\delta<1$, its most 
likely value being $\delta\simeq\,0.395$. 
\\
Let us now study the implications of the above results for $c^2$ and $b^2$. 
Using Eq.(\ref{pri}) and the constraint (\ref{rrbc}) we obtain
\\
\be c^2+b^2=\frac{r^{-}}{1+r^{-}}\left[\frac{1-2q^{-}}{3}\right]\simeq
\frac{r_{0}}{1+r_{0}}\left[\frac{1+2\delta}{3}\right]\,\simeq\,0.179.
\en
\\
To assign a  value to $r^{+}$ (which may be identified with the value of $r$
at the end of domain walls formation), we assume it should not be far from  the 
$r$ value at the epoch of primordial nucleosynthesis, $r_{N}$. Note that the 
nucleosynthesis epoch is the earliest one on which we have reliable quantitative 
information 
and that the time elapsed between domain walls formation and nucleosynthesis 
is larger by many magnitude orders than the time elapsed between 
the latter and the present epoch. As can be checked 
(see the Appendix of Ref.\cite{ladw}),   
\\
\be
r^{+}\,\simeq\,r_{N}\,>\frac{4\,g_{\ast}(T_{N})}{7\Delta\,N_{max}}\simeq\,6.143,
\en
\\
where the maximum variation in the effective number of neutrino
species satisfies  $\Delta\,g_{\ast}(T_N)<7\Delta\,N_{max}/4$,
with $\Delta\,N_{max}=1$ and $g_{\ast}(T_N)=10.75$ for the 
standard model with three massless neutrinos \cite{dolgov}. 
Assuming $r^{+}\simeq\,r_{N}$ and
$r^{-}\simeq\,r_{0}=3/7$ in Eq.(\ref{pri}), it follows that
$c^{2}<0.068$. Accordingly, $b^{2}$ must lie in the range
0.111$<b^2<$ 0.179, whereby $0<c^2<0.068$.
From the constraint (\ref{rrbc}) we obtain
\\
\be 
\frac{b^{2}}{c^{2}}\simeq\,r_N\,r_{0}-1\,>1.633, 
\label{bc}
\en
\\
suggesting that the viscous effects are rather important.
Note that because the time of domain walls formation 
precedes primordial nucleosynthesis and $r$ decreases with expansion,
$r_{N}$ just represents a lower bound on the value of $r$ 
at that time.

\section{Statefinder parameters}
Due to the ample choice of dark energy models the
deceleration parameter is unable by itself to discriminate
between them nor even from the $\Lambda$CDM model. This is why
Sahni {\it et al.} \cite{Sahni} introduced two dimensionless,
geometrical parameters, the so--called statefinder pair
$(\ov{r},\ov{s})$, to more fully characterize the different models
and in particular to help discriminate between the $\Lambda$CDM
model and those based on a varying equation of state. These
parameters are defined as
\\
\be
\ov{r} =\frac{\dddot{a}}{aH^{3}}, \qquad
\ov{s}=\frac{\ov{r}-1}{3(q-\frac{1}{2})}  \, .
\en
\\

In our case, we find that at late times
\be
\ds \, q=-\left(1-\frac{\beta}{2}\right), \quad
\ov{r}=1-\frac{3\beta}{2}+\frac{\beta^2}{2}, \quad \ov{s} =\frac{\beta}{3},
\label{qrs}
\en
\\
being related by $\ov{r} = \,1\,+\,\textstyle{9\over{2}} \ov{s}(\ov{s}-1)$.

In more general situations such that $w$ may not be reasonably
well approximated by a constant, the deceleration parameter and
the statefinder pair can be calculated from the Friedmann equation
(\ref{Friedmann}) and the evolution equation for the Hubble factor
given by Eq.(\ref{dotH}). The result is now
\\
\be
q = \textstyle{1\over{2}} \left[1 + 3(w \Omega_{dw} - b^{2}) \right],
\label{q2}
\en
\\
\begin{eqnarray}
\ov{r}=&1&-\frac{9\,\Omega_{dw}
}{2}\left\{\phantom{\frac{.}{.}}(1-b^2)[b^2(r+1)+1]-(1+w)[(1+w)-b^2]\right. \\
\nonumber
       &+&\left.\frac{b^{2}w}{r}-\frac{rw^{2}}{r+1}-w-\frac{\dot{w}}{3H}\right\},
\label{r2}
\end{eqnarray}
and
\\
\begin{eqnarray}
\ov{s}&=&\frac{\Omega_{dw}}{(b^2-w\,\Omega_{dw})}\left\{\phantom{\frac{.}{.}}(1-b^2)[b^2(r+1)+1]-(1+w)[(1+w)-b^{2}]\right. \\
\nonumber
      &+&\left. \frac{b^{2}w}{r}+
      \frac{rw^2}{r+1}+w+\frac{\dot{w}}{3H}\right\},
\label{s2}
\end{eqnarray}
where $\Omega_{dw} \equiv \kappa\,\rho_{dw}/(3H^{2})$ is the density parameter 
of the domain walls component.

\section{Discussion}
We have presented a cosmological model of late
acceleration driven by domain walls coupled with dissipative dark
matter fully consistent with the high redshift supernovae
data. At late times the coupling leads to a stable constant ratio between the
energy densities of both components compatible with accelerated
expansion, thus solving the coincidence problem without recourse to
fine tuning.  This ratio can be thought as constant at late times only,
thereby we have studied its evolution with the Universe expansion and 
estimated its value, $r_{N}$, at the nucleosynthesis epoch. This sets
a lower bound on the value of $r$ at the time of domain walls formation.
In so doing we have implicitly assumed that the interaction domain 
walls-dark matter starts right after the former component comes into existence.
In a more detailed analysis the radiation component should enter  the 
picture as well as it dominates the expansion from the end of the inflationary
era till matter-radiation equality. However, this phase  is extremely short 
compared with the matter plus dark energy dominance period. We believe 
its influence should not substantially alter the main 
conclusions of the paper. 

We have, moreover,  calculated the statefinder parameters of this model.  
It is to be hoped that in some not distant future observational 
techniques able to measure the $(\ov{r},\ov{s})$ pair will be 
in place so that these two parameters will become useful
tools to ascertain the nature of dark energy.

\acknowledgments
SdC thanks T. Vachaspati for useful comments. RH wishes to thank
Universidad Aut\'onoma de Barcelona  for its hospitality where
part of this work was done. SdC was supported by the Comision
Nacional de Ciencia y Tecnolog\'{\i}a through FONDECYT (Grants
$1030469$, $1010485$ and 1040624) and by PUCV (Grant
$123.764/2004$). RH was supported from Ministerio de Educaci\'on
through MECESUP projects FSM 9901 and from PUCV through Proyecto
de Investigadores J\'ovenes a$\tilde{n}$o 2004. DP acknowledged
the support from the Spanish Ministry of Science and Technology
(grant BFM2003--06033).

\end{document}